\documentclass{emulateapj}

\usepackage{graphicx}
\usepackage{dcolumn}
\usepackage{bm}

\usepackage{amsmath}
\usepackage{amssymb}

\newcommand{\ep}{\epsilon}
\newcommand{\mm}{\mu_{max}}

\newcommand{\jcap}{JCAP}
\newcommand{\na}{New Astronomy}

\begin{document}

\title{The galaxy halo formation in the absence of violent relaxation and a universal density profile of the halo center.}

\author{A. N. Baushev}
 \affil{DESY, 15738 Zeuthen, Germany}
 \affil{Institut f\"ur Physik und Astronomie, Universit\"at Potsdam, 14476 Potsdam-Golm, Germany}
 \email{baushev@gmail.com}

\begin{abstract}
While N-body simulations testify for a cuspy profile of the central region of the dark matter
haloes, observations favor a shallow, cored density profile of the central region of, at least,
some spiral galaxies and dwarf spheroidals. We show that a central profile, very close to the
observed one, inevitably forms in the center of dark matter haloes if we make a supposition about a
moderate energy relaxation of the system during the halo formation. If we assume the energy
exchange between dark matter particles during the halo collapse to be not too intensive, the
profile is universal: it depends almost not at all on the properties of the initial perturbation
and is very akin, but not identical, to the Einasto profile with small Einasto index $n\sim 0.5$.
We estimate the size of the 'central core' of the distribution, i.e., the extent of the very
central region with a respectively gentle profile, and show that the cusp formation is unlikely,
even if the dark matter is cold. The obtained profile is in a good agreement with observational
data for, at least, some types of galaxies, but clearly disagrees with N-body simulations.
\end{abstract}


\section{Introduction}
The problem of the Universe structure formation is still far from the complete solution. Since the
task is essentially nonlinear, N-body simulation is one of the possible ways to solve it. While
their results satisfactorily describe observed properties of galaxy clusters \citep{okabe2010},
some discrepancy seems to appear when we consider galaxy haloes. N-body simulations predict a very
steep, NFW-like density profile of the central region of the dark matter haloes \citep{nfw}. The
Navarro-Frenk-White profile behaves there as $\rho\propto r^{-1}$. Though recent simulations favour
the Einasto profile \citep{mo98, gao2008, diemand2008, mo09, navarro2010}
\begin{equation}
\rho=\rho_s \exp\left[-2 n\left\{\left( \frac{r}{r_s}\right)^\frac{1}{n} -1\right\}\right]
\label{15a0}
\end{equation}
with a finite density in the center \citep{einasto}, the index of the profile is so high
(typically, $n\simeq 5-6$) that it can also be considered as a cuspy one.

On the contrary, observations  \citep{deblok2001, bosma2002, marchesini2002, gentile2007} show that
galaxies have a core (i.e. a region with rather a shallow density profile) in the center.
\citet{mamon2011} fitted the dark matter density profiles of a large array of spiral galaxies by
the Einasto profile, considering the Einasto index $n$ as a free parameter. They found that Einasto
profile provides a significantly better fit of observational data of the central region than either
NFW or the cored pseudo-isothermal profiles. However, the Einasto index is found to be small
($n\simeq 0.5-1$), that actually corresponds to a cored profile and is not in agreement with the
predictions from $\Lambda$CDM simulations. Dwarf galaxies also show no cusps in their centra
\citep{oh2011, governato2012}.

There are several possible ways to account for this disagreement. It can be caused by the influence
of baryon matter \citep{pontzen2012}. However, this explanation is not doubtless. The fraction of
baryons in some dwarf galaxies is so tiny that it hardly can influence the cusp formation
\citep{garrison-kimmel2013}, but these objects show no cusps in their centra. Meanwhile, galaxy
clusters have a much larger fraction of baryons, and their density profiles are rather steep
\citep{okabe2010}. The problem of the baryon influence on the dark matter distribution is very
complex and yet not quite clear. It lies beyond the scope of the present work, and we will not
discuss it in this paper anymore, addressing the readers to the vast literature on this topic (see
\citep{weinberg2013, pontzen2014} and references therein). The absence of the cusps could be a sign
that the dark matter is warm and consists of light particles like sterile neutrinos. Thus the 'cusp
vs. core' problem might shed light upon the physical nature of dark matter. However, we should
first be sure that there is no explanation to the cores in the framework of cold DM paradigm.


We will try to show by this paper that the energy evolution of the forming halo is probably the key
to the problem of central cusp occurrence. Though the density shape in the center of the halo may
differ from the NFW profile, we will use the NFW halo concentration $c_{vir}$ to qualitatively
characterize a halo, because of popularity of the NFW profile. Let us denote by
$\ep=\frac{v^2}{2}+\phi$ the total energy of a unit mass of the matter, $\phi$ is the gravitational
potential. It is easy to show (see the beginning of section \ref{sec2}) that in the very general
case the particles that later form the halo have very narrow initial energy distribution: the
initial energies of the particles differ no more than $2$ times. When a halo collapses, remarkable
inhomogeneities and caustics appear; their strong small-scale gravitational field mediates a
relaxation and may significantly smear out the energy spectrum of the dark matter particles with
respect to the initial one. A question appears: how strong can the relaxation be, i.e., can the
energy evolution of the system be arbitrarily strong, or the ratio between the final $\ep_f$ and
the initial $\ep_i$ energies is somehow limited for the majority of the particles?

In principle, the ratio $\ep_f/\ep_i$ could be arbitrarily large: the idea of the 'violent
relaxation' \citep{violent} was the supposition that a particle completely forgets its initial
state during the relaxation, and its final energy has no connection with the initial one. However,
the equilibrium solution obtained by \citet{violent} has infinite total mass, and its applicability
to the finite systems is discussible. The main fraction of mass of the future halo is initially
concentrated in the outer layers with the highest initial $r$ (since their volume grows as $r^2$).
Meanwhile, as it was shown by \citet{violent}, the efficiency of all relaxation processes (and so
the alternation of the particle energy) rapidly drops with $r$: even in that work the outer part of
the system remained unrelaxed. Moreover, simulations show that a significant part of the total mass
of the halo accretes to the already collapsed halo \citep{wang2012}. This substance has a
respectively high total energy and cannot be significantly affected by the violent relaxation.

Unfortunately, the halo collapse is a very complex nonlinear process, and there is almost no hope
to find an exact theoretical solution of the problem. The literature offers several possible
approaches to the task. \citet{syer1998} suggested that the cuspy DM haloes might arise as a
consequence of hierarchical structure formation. However, this theory predicts very steep cusps
($\rho\propto r^{-2}$) in the case of Zeldovich-Harrison spectrum of initial perturbations. This
cusp shape is hardly comparable with modern observations and simulations, while \citet{planck}
firmly confirms the Zeldovich-Harrison spectrum. An another approach is to introduce Boltzmann's
H-function \citep{tremaine1986, stiavelli1987} that is actually the entropy of the system. The
system cannot reach the real maximum of entropy, which is the isothermal sphere, but the authors
suggest that the H-function increasing defines the direction of the system evolution. However,
\citet{pontzen2013} correctly pointed to the possible difficulties of this approach. The violent
relaxation (whatever violent it is) is actually a collisionless relaxation through the common
field. This process is, in principle, reversible, and the entropy does not grow in it \citep{ll5,
pontzen2013}. A good illustration for this statement is the existence of stationary solutions, like
\citet{osipkov1979} or \citet{jaffe1983}. These systems are very far from the entropy maximum, in
particular, the local velocity distribution there is not Maxwell. However, they cannot relax
anymore, since their gravitational field is already stationary. The only process that guarantees
the entropy growth is the particle collisions. The characteristic of the collisional relaxation is
\citep[eqn. 1.38]{bt}
\begin{equation}
\tau_r= \dfrac{N(r)}{8 \ln\Lambda}\tau_d
 \label{15c1}
\end{equation}
where $v$, $N(r)$, $\ln\Lambda$, and $\tau_d=r/v$ are the 'characteristic' particle velocity, the
number of particles inside radius $r$, the Coulomb logarithm, and the dynamical time of the system
at $r$, respectively. The dynamical time of the galaxies is typically only a few orders of
magnitude smaller than the age of the Universe. Since a galactic halo may contain $\sim 10^{65}$
particles (accepting the DM particle mass $200$~{GeV}), the relaxation time enormously exceeds the
age of the galaxies. So the entropy growth can be completely negligible at the cosmological time.
Moreover, the system can hardly be ergodic in this case, which is necessary to use the statistical
methods \citep{pontzen2013}. Surely, the statistically stable state is the isothermal sphere, i.e.
it is cuspy. However, the applicability of the essentially statistical approach to a system that
has existed for only a tiny fraction of the true relaxation time is discussable.

N-body simulations give a more direct way to investigate the halo relaxation. The idea of the
method is to substitute real tiny DM particles by massive test bodies with smoothed Newtonian
potential. Recent N-body simulations \citep{diemand2005, diemand2007, diemandkuhlen2008} indicate
that the supposition of violent relaxation is not true: the energies of the majority of the
particles do not change completely and correlate with the initial ones. Nevertheless, the
relaxation is rather strong, and the halo profiles are cuspy.

The method has a weak point, however: the number of the test bodies is small as compared with the
real DM particles, and the influence of the unphysical test body collisions should be carefully
avoided. According to (\ref{15c1}), the relaxation time is proportional to $N(r)$. The closer we
approach the halo center, the smaller the number of the bodies $N(r)$ is. At some radius $r_{conv}$
the ratio of the simulation time $t_0$ to the relaxation time $\tau_r$ may go so large that the
collisions become important. So the simulations are still reliable outside the convergence radius
$r_{conv}$, but the density profile is already corrupted by the unphysical test body collisions
inside $r_{conv}$.

As we have already mentioned, the dark matter in simulations rather rapidly forms stationary haloes
with an almost universal density profile. The profile is close to the NFW, being cuspy
($\rho\propto r^{-\gamma}$, $\gamma \simeq 1$) or very steep in the center. The persist appearance
of the cusp and its stability with respect to parameter variations of N-body codes is usually
considered as a proof of its physical nature and independence on the collisions. The commonly-used
method \citep{navarro2010} to assure the negligibility of the the collisions is to find the moment
when the the cusp starts to smear out, which is believed to be the the first sign of the collision
influence \citep{power2003}. However, the cusp turns out to be surprisingly stable.
\citet{power2003} found that the smearing does not appear at least up to $t=1.7 \tau_r$ and
probably much longer. Indeed, \citet{hayashi2003, klypin2013} observed that the cusp was stabile
even at tens of relaxation times, and only then smeared out. The negligibility of the test body
collisions at $t\sim 30 \tau_r$ seems suspicious. In order to test the influence of the particle
encounters analytically the Fokker-Planck equation was used \citep{evans1997, 13}. It turned out
that the density distribution close to NFW ($\rho\propto r^{-\gamma}$, $\gamma \simeq 1$) is an
attractor solution for the kinetic equation, taking into account close encounters. The
Fokker-Planck diffusion occurring as a result of the collisions transforms any initial density
distribution shallower than $\rho\propto r^{-1}$ into a NFW-like profile ($\gamma \simeq 1$) in a
time $t<\tau_r$. Being formed, the cuspy profile should survive for a much longer time interval,
since the collision effects are self-compensated to a first approximation in this case. Moreover,
since the shape of the attractor solution is not sensitive to the simulation parameters, the cusp
should be insensitive as well. At tens of relaxation times $\tau_r$ the NFW-like profile should be
smeared out by the 'thermal conductivity', which is a second-order effect of particle collisions
\citep{quinlan1996, 13}.

These results perfectly describe the simulation behavior. The cusp forms at $t<\tau_r$, survives
for several or even many relaxation times, and then smears out in tens relaxation times. However,
since the test body collisions are unphysical and correspond to nothing real, the cusp formation by
the Fokker-Planck diffusion is a purely numerical effect. As we could see, the collisions of real
DM particles are completely ineffective. Thus the commonly-used criterion of the convergence radius
$r_{conv}$ of simulations $t<1.7 \tau_r$ \citep{power2003} seems excessively optimistic. The fact
that the cusp is stable and insensitive to the simulation parameters is not enough to state that
the profile is not affected by the unphysical collisions. The only reliable criterion is $t\ll
\tau_r$. However, it claims much more test bodies inside $r_{conv}$ and so increases several times
the convergence radius. This can be of vital importance for the 'cusp vs. core' problem. Actually,
the disagreement between the simulations and observations occur only quite close (a few percent of
$R_{vir}$) to the halo center. If we use the more reliable criterion $t\ll \tau_r$, the
contradictions may disappear.

\begin{figure}
 \resizebox{\hsize}{!}{\includegraphics[angle=0]{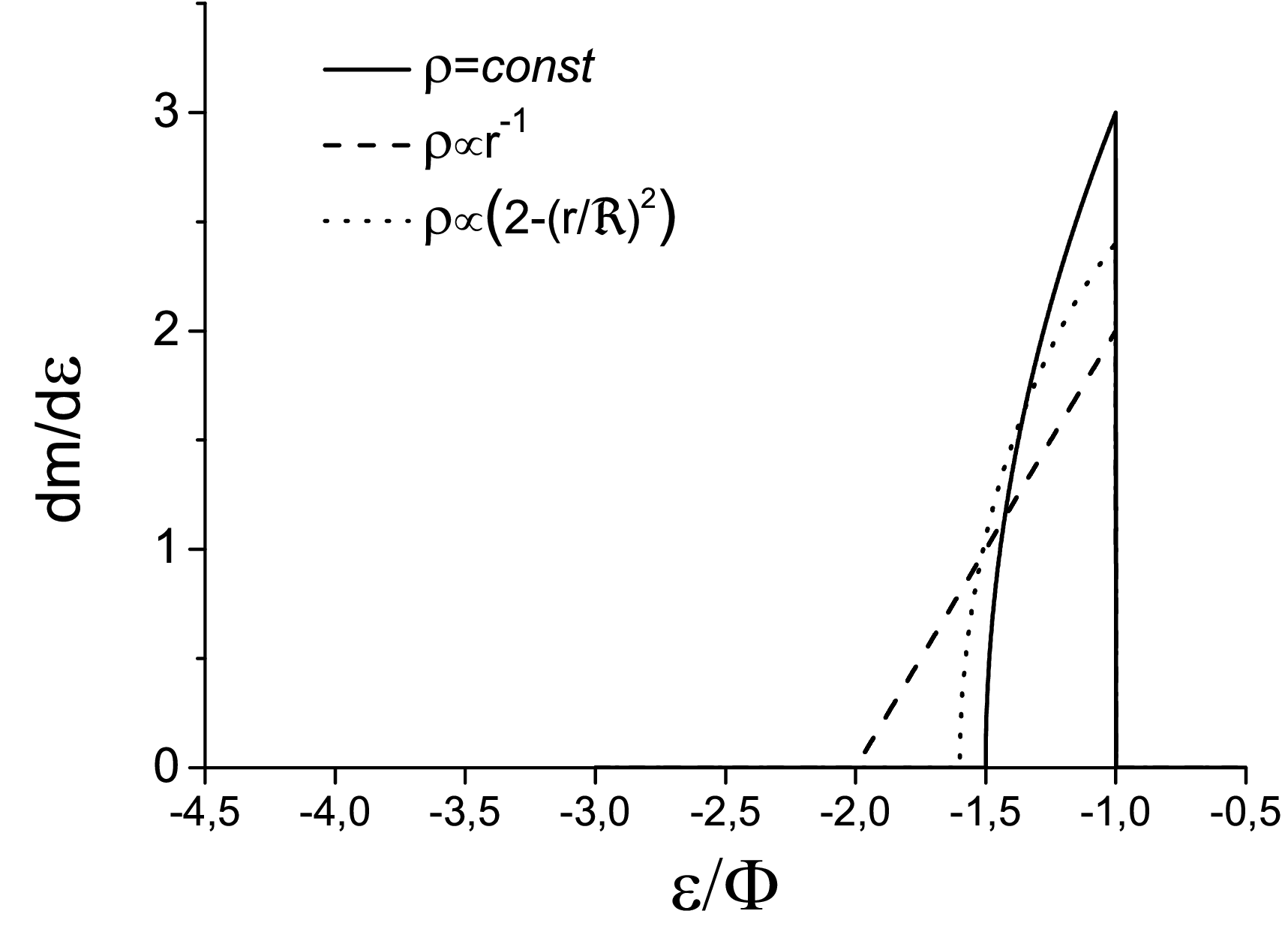}}
\caption{Energy spectra $\frac{dm}{d\ep}$ for various forms of initial perturbations. Apparently,
there are no particles with $\ep>-\Phi$.}
 \label{15fig1}
\end{figure}

The main aim of this paper is to show that a cored profile, very similar to that observed in the
centra of the galaxies, inevitably forms in the center of dark matter haloes if we make the only
supposition that the energy relaxation of the galactic dark matter haloes was \emph{moderate} in
the following sense:

\emph{The principle assumption.} For the majority of the particles, the final energy in the formed
halo $\ep_f$ differs no more than $c_{vir}/3$ times from the initial value $\ep_i$
\begin{equation}
\label{15a1} \frac{\ep_f}{\ep_i}\le \frac{c_{vir}}{3}
\end{equation}

Concentration $c_{vir}$ of the Milky Way galaxy lies between $12$ and $17$ \citep{klypin}, and
should be even higher for smaller galaxies \citep{navarro2010}. So the moderate relaxation
supposition means that the energies of the majority of the particles have not changed more than
$4$-$5$ times for the Milky Way galaxy; for less massive haloes the assumption is even softer. Of
course, there are always particles, which energies have changed more than (\ref{15a1}); however, we
suppose that their fraction is small. The condition of the moderate relaxation will be specified
and restated in the Discussion section.

We should underline that we suggest the moderate relaxation only for relatively low-mass objects
like haloes of galaxies. Moreover, despite of all the reasoning above, this is no more than a
hypothesis: now the assumption cannot be certainly proved, while the validity of all results of
this article depends on the feasibility of this basic supposition. However, as we will see, the
fulfilment of condition (\ref{15a1}) automatically leads to a central density profile in good
agreement with observations. For this reason alone it is worthy to consider the moderate
relaxation, at least as a hypothesis.

\begin{figure}
\resizebox{\hsize}{!}{\includegraphics[angle=0]{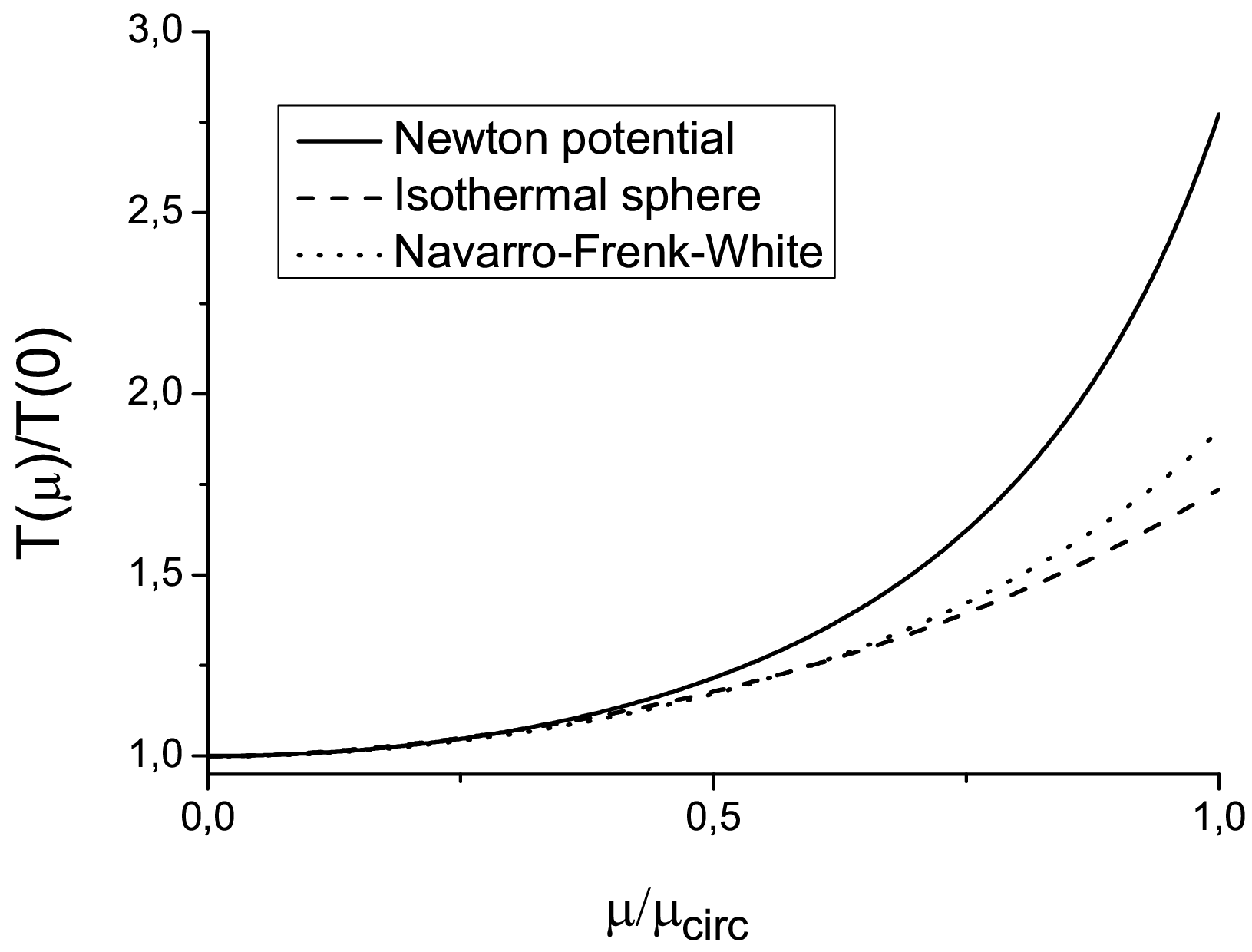}}
 \caption{Ratios $T(\mu)/T(0)$ for the newtonian potential $\phi\propto -1/r$ (solid line), isothermal
 sphere $\phi\propto ln(r)$ (dashed line) and Navarro-Frenk-White profile with $c_{vir}=5$ (dotted line).
 The normalization of the potentials is chosen so that the mass inside $r_0$ is the same for all three
 cases. $\mu_{circ}$ corresponds to the circular orbit.}
 \label{15fig2}
\end{figure}

\section{Energy spectrum}
\label{sec2}

 Let us consider the initial distribution of matter. Hereafter in this article we will
consider a spherically-symmetric task and neglect the presence of baryon matter. The mass and
initial radius of the halo are $m$ and $\mathfrak R$, $\mathfrak R\simeq R_{vir}$. If the dark
matter is cold, the initial velocity of the matter may be thought of as to be zero without loss of
generality \citep{gorbrub2}. (As we will see at the end, the conclusions of this section is also
valid for the warm dark matter). To begin with, let us consider the case when the matter is
uniformly distributed inside $\mathfrak R$ ($\rho=\it{const}$, the Tolman case). Inside the sphere:
\begin{equation}
\label{15a2}
 \phi(r) = \frac{Gm}{\mathfrak R}\left(\frac12 \left(\frac{r}{\mathfrak R}\right)^2 - \frac32\right)
\end{equation}
\begin{equation}
\label{15a3}
 dm = 3 \frac{m}{\mathfrak R} \left(\frac{r}{\mathfrak R}\right)^2 dr
\end{equation}
Hence we obtain the initial energy distribution of the substance:
\begin{equation}
\label{15a4}
 \frac{dm}{d\ep} =  \left(\frac{3\mathfrak R}{G}\right) \sqrt{3+ \frac{2\ep}{\Phi}};\qquad
\ep\in [-\frac32 \Phi;-\Phi]
\end{equation}
where $\Phi\equiv\frac{Gm}{R_{vir}}\simeq\frac{Gm}{\mathfrak R}$. Performing similar calculations,
we obtain for the case of initial density perturbation $\rho\propto r^{-1}$
 \begin{equation}
 \label{15a5}
 \frac{dm}{d\ep} = \frac{2m}{\Phi} \left(2+\frac{\ep}{\Phi}\right);\qquad \ep\in [-2 \Phi;-\Phi]
\end{equation}
As one can see in Fig.~\ref{15fig1}, distributions (\ref{15a4}) and (\ref{15a5}) differ not so
strongly: all the particles lie in a respectively narrow energy range $[-2\Phi;-\Phi]$ and the
majority of the particles have energies close to $-\Phi$. Moreover, the initial energy spectrum of
any real (smooth) density perturbation lies between (\ref{15a4}) and (\ref{15a5}): the perturbation
collapses when $\frac{\delta\rho}{\rho}= \beta\simeq 1$, and $\beta$ weakly depends on its shape.
Consequently, if the initial shape of $\delta\rho(r)$ is a smooth function, distributions of $\phi$
and $dm$ differs from (\ref{15a2}) and (\ref{15a3}) only on a multiplier of the order of $2$. As a
result, distribution $dm/d\ep$ does not differ very significantly from the uniform one
(\ref{15a4}). As an illustration, we also consider profile  $\rho\propto (2-(r/\mathfrak R)^2)$
that well approximates real initial perturbations. The corresponding energy distribution,
represented in in Fig.~\ref{15fig1}, turns out to be quite close to (\ref{15a4}).

It seems reasonable to say that the initial energy distribution is similar to (\ref{15a4}) under
any reasonable choice of shape of the initial perturbations. If we divide a spherically symmetric
initial distribution on radial layers of thickness $dr$, $\ep$ is constant within the layer, and
the volume of the layer grows as $\left(\frac{r}{\mathfrak R}\right)^2$. As a result, the particles
with $\ep\simeq -\Phi$ (which corresponds to $r=\mathfrak R$) dominate in the initial energy
spectrum even in the case of quite a steep profile ($\delta \rho\propto r^{-1}$). A strong
departure from (\ref{15a4}) will take place only if the initial profile is very steep (for
instance, $\delta \rho\propto r^{-3}$). On the other hand, such profiles are hardly suitable to
describe the initial perturbations of dark matter.

So for a large variety of initial conditions the initial energy distribution of the particles is
similar to (\ref{15a4}), at least, in the following sense: the majority of the particles have the
energy close to $-\Phi$, and their energy distribution is quite narrow. In case of distribution
(\ref{15a4}), for instance, for all the particles $\ep\in [-\frac32\Phi;-\Phi]$. This conclusion is
also valid for the warm dark matter. Indeed, the smearing of the energy spectrum with respect to
the cold dark matter case is of the order of the temperature of dark matter at the moment of the
structure collapse. For the structure to be able to collapse, the temperature should be much lower
than $\Phi$.

In view of our principle assumption that the energy distribution of the particles changes
moderately with respect to the initial one, the similarity between the initial distributions leads
to a similarity of the particle energy spectra of the relaxed halo. But how can the system evolve
at all, if the energies of the bulk of the particles remain almost the same? The mechanism is the
following: the particles oscillate in the potential well. Initially they oscillate cophasal. Then,
even if energies of the particles remain constant, their phases diverge. Typically, the period
sharply depends on $\ep$ for the majority of potential wells. But even if the energies of the
particles precisely coincide, their phases can diverge as a result of a small nonsphericalness,
perturbations from the nearby structures etc. Finally, the phases of the oscillations are mixed
thoroughly, and each phase becomes equiprobable. This is exactly the situation, corresponding to
the stationary halo. It is important to underline that the phase mixing goes on continuously, in
contrast to the energy exchange, which is effective only during the first short stage of violent
relaxation.

Finally, the halo collapse forms a stationary potential well $\phi(r)$, where $\phi$ does not
depend on time. Each particle can be characterized by its specific angular momentum $\mu= |[\vec
v\times\vec r]|$ and maximum radius $r_0$ from the center the particle can reach. Radius $r_0$ is
one-to-one bound with the specific particle energy by evident relationship $\ep=\phi(r_0)+\mu^2/(2
r_0^2)$. We may consider the halo particle distribution function over $r_0$.
 \begin{equation}
 \label{15a6}
dm = f(r_0)\ dr_0
\end{equation}

If the energy evolution of the halo is moderate, $f(r_0)$ has a very peculiar appearance. Indeed,
the potential well of the collapsed halo is much deeper than the initial one. An initial
perturbation had $k\equiv\left|(\phi(R_{vir})-\phi(0))/\phi(R_{vir})-\phi(\infty)))\right|\le 1$,
as we could see from eqn.~\ref{15a4}-\ref{15a5}, while a formed galactic halo always has $k\gg 1$.
For instance, a Navarro-Frenk-White halo has $k\simeq c_{vir}$ \citep{nfw, 9}. Accepting for the
Milky Way $M_{vir}=10^{12} M_\odot$, $R_{vir}=250$~{kpc} \citep{klypin}, we obtain
$-\Phi=-\phi(R_{vir})\simeq (130 \text{km/s})^2$. Meanwhile, escape speed near the Solar System
undoubtedly exceeds $525$~{km/s} \citep{escape} and may in principle be much larger ($650$~{km/s}
or even higher \citep{suchkov, bt}). It means that $k\ge 10$ for the Milky Way and should be even
higher for less massive systems. The specific energies of the particles bound in such a well may in
principle lie between $\ep=-\Phi$ ($r_0\simeq R_{vir}$) and $\ep=-k \Phi\simeq -c_{vir}\Phi$
($r_0=0$). However, if we accept \emph{principle assumption} (\ref{15a1}), the real energy range
occupied by particles is narrower. Indeed, as we could see, the particles are strongly concentrated
towards $\ep=-\Phi$ in any reasonable initial perturbation, and if we consider profile $\rho\propto
(2-(r/\mathfrak R)^2)$ as a good approximation of real initial perturbations, we saw that the
energies of all the particles fall in $\ep\in [-\Phi;-1.6\Phi]$ (see Fig.~\ref{15fig1}). If
assumption (\ref{15a1}) is true, it suggests that the majority of the particles still concentrate
near $\ep=-\Phi$: the energy exchange of the particles during the relaxation is more or less a
stochastic process, while the total energy of the system should conserve. Thus the narrowness of
the initial energy distribution, the relative smallness of the energy evolution, and the fact that
the final potential well is much deeper, than the initial one, together result in crowding of the
particle apocentre distances $r_0$ near $R_{vir}$.

However, a much more important for us consequence of assumption (\ref{15a1}) is that the halo
contains almost not at all particles with energies less than $\ep=-1.6\Phi\times
\dfrac{c_{vir}}{3}\simeq -\dfrac{c_{vir}}{2} \Phi$. Radius $r_0$, corresponding to $\ep=
-\dfrac{c_{vir}}{2} \Phi$, depends on the density profile. However, it a fortiori
exceeds\footnote{Here we neglect the part of the particle energy associated with the angular
momentum $\dfrac{\mu^2}{2 r^2}$. N-body simulations show that the orbits of the majority of the
particles are rather prolate \citep{diemandkuhlen2008}, and in this case the influence of the
angular momentum on $r_0$ is negligible. However, even in the worse case of circular orbit, $r_0$
of a particle cannot be more than two times smaller than one for a radially moving particle with
the same energy.} $2 \frac{R_{vir}}{c_{vir}}$, which corresponds to the gravitational field of a
point mass $M_{vir}$; for a high-concentrated NFW halo $\ep= -\dfrac{c_{vir}}{2} \Phi$ corresponds
to $r_0\simeq 2.5 \frac{R_{vir}}{c_{vir}}$. Since condition (\ref{15a1}) states that a halo
contains only a few particles with $\ep< -\dfrac{c_{vir}}{2} \Phi$, it means that the halo contains
only a few particles with $r_0\in[0; 2 \frac{R_{vir}}{c_{vir}}]$. Actually, a number of particles
with $r_0\sim 2 \frac{R_{vir}}{c_{vir}}$ should also be relatively small, since initial
distribution $\rho\propto (2-(r/\mathfrak R)^2)$ contains only a few particles with $\ep\simeq -1.6
\Phi$. We may conclude that distribution function $f(r_0)$ has a steep maximum near $r_0\sim
R_{vir}$ and that it is almost equal to zero when $r_0\sim (2-3) \frac{R_{vir}}{c_{vir}}$.

The later property is the most important for us. Though the fraction of particles with $r_0\le
(2-3) \frac{R_{vir}}{c_{vir}}$ is small, the region $r\le (2-3) \frac{R_{vir}}{c_{vir}}$ is quite
large ($40$-$50$~{kpc} for our Galaxy) and contains a very significant fraction of the galaxy mass.
It means that the dominant contribution into the dark matter profile in the halo center is given by
particles that come there from the outside, i.e. that have apocenter distances $r_0$ significantly
larger than the radius of the region under discussion. Hereafter we will consider even smaller area
around the halo center, $r\le \frac{R_{vir}}{c_{vir}}$, where we thus have $r\ll r_0$ for almost
all the particles. On the other hand, this area is still quite large
($\frac{R_{vir}}{c_{vir}}\simeq 20$~{kpc} for the Milky Way galaxy). Moreover, $r=
\frac{R_{vir}}{c_{vir}}$ corresponds to the radius where a NFW halo has $\frac{d\log\rho(r)}{d\log
r}= -2$; consequently, we may expect that if a halo has a core, its radius is smaller than the
radius of the area under consideration.

As we will see, if the central profile is created by particles with $r\ll r_0$, the dark matter
density profile is quite universal, does not depend on the detailed properties of distribution
$f(r_0)$ and is close to the Einasto profile with $n\simeq 0.5$.

\section{Calculations}
We can easily find the density distribution in the center of the halo, created by the particles
coming to this region from the outside, using the method offered in \citep{14}. Let us single out
the particles with a certain $r_0$. We denote their total mass by $m$. N-body simulations
\citep{2006JCAP...01..014H, kuhlen2010} suggest that the distribution over tangential velocities is
close, but do not coincide with Gaussian. According to these results, we assume for simplicity that
specific angular momentum $\mu\equiv[\vec v\times\vec r]$ of the particles has Gaussian
distribution
 \begin{equation}
 \label{15a7}
dm = m\frac{2\mu}{\alpha^2} \exp\left(-\frac{\mu^2}{\alpha^2}\right)\; d\mu
\end{equation}
Here $\alpha\equiv \alpha(r_0)$ is the width of the distribution; generally speaking, $\alpha$ is a
function of $r_0$. As we could see, the majority of the particles have $r_0\sim R_{vir}$, and so
only those with small $\mu$ can penetrate into the area of our interest $r\sim R_{vir}/c_{vir}$. So
the distribution in the halo center is mainly determined by the behavior of (\ref{15a7}) when
$\mu\to 0$. Thus our calculation is not very sensitive to the assumption of Gaussian distribution:
any other distribution with the same behavior at $\mu \simeq 0$ would give a similar result.

 A particle moving in gravitational field $\phi(r)$ has two integrals of motion:
$\mu$ and $\ep=\dfrac{v_r^2}{2}+\dfrac{\mu^2}{2 r^2}+\phi$. The radial velocity of the particle is
equal to
 \begin{equation}
 \label{15a8}
 v_r = \sqrt{2(\phi(r_0)-\phi(r))-\mu^2\left(\frac{1}{r^2}-\frac{1}{r_0^2}\right)}
\end{equation}
We introduce the maximum angular momentum of a particle wherewith it can reach radius $r$
 \begin{equation}
 \label{15a9}
 \mm^2 = 2 (\phi(r_0)-\phi(r)) \left(\frac{1}{r^2}-\frac{1}{r_0^2}\right)^{-1}
\end{equation}
Then (\ref{15a8}) may be rewritten as
 \begin{equation}
 \label{15a10}
 v_r = \frac{\sqrt{r_0^2-r^2}}{r r_0} \:\sqrt{\mm^2-\mu^2}
\end{equation}
A particle having maximal radius $r_0$ and minimal radius $r_{min}$ contributes into the halo
density on all the interval $[r_{min}, r_0]$. The contribution of a single particle of mass $m$ on
an interval $dr$ is proportional to the time the particle passes on this interval \citep{11}.
 \begin{equation}
 \label{15a11}
 \frac{dm}{m} = \frac{dt}{T} = \frac{dr}{v_r T}
\end{equation}
Here $T$ is the half-period of the particle, i.e. the time it takes for the particle to fall from
its maximal radius to the minimal one
 \begin{equation}
 \label{15a12}
 T(r_0,\mu) = \int^{r_0}_{r_{min}}\;\frac{dr}{v_r}
\end{equation}
$T(r_0,\mu)$ is, generally speaking, a function of $r_0$ and $\mu$.

\begin{figure}
 \resizebox{\hsize}{!}{\includegraphics[angle=0]{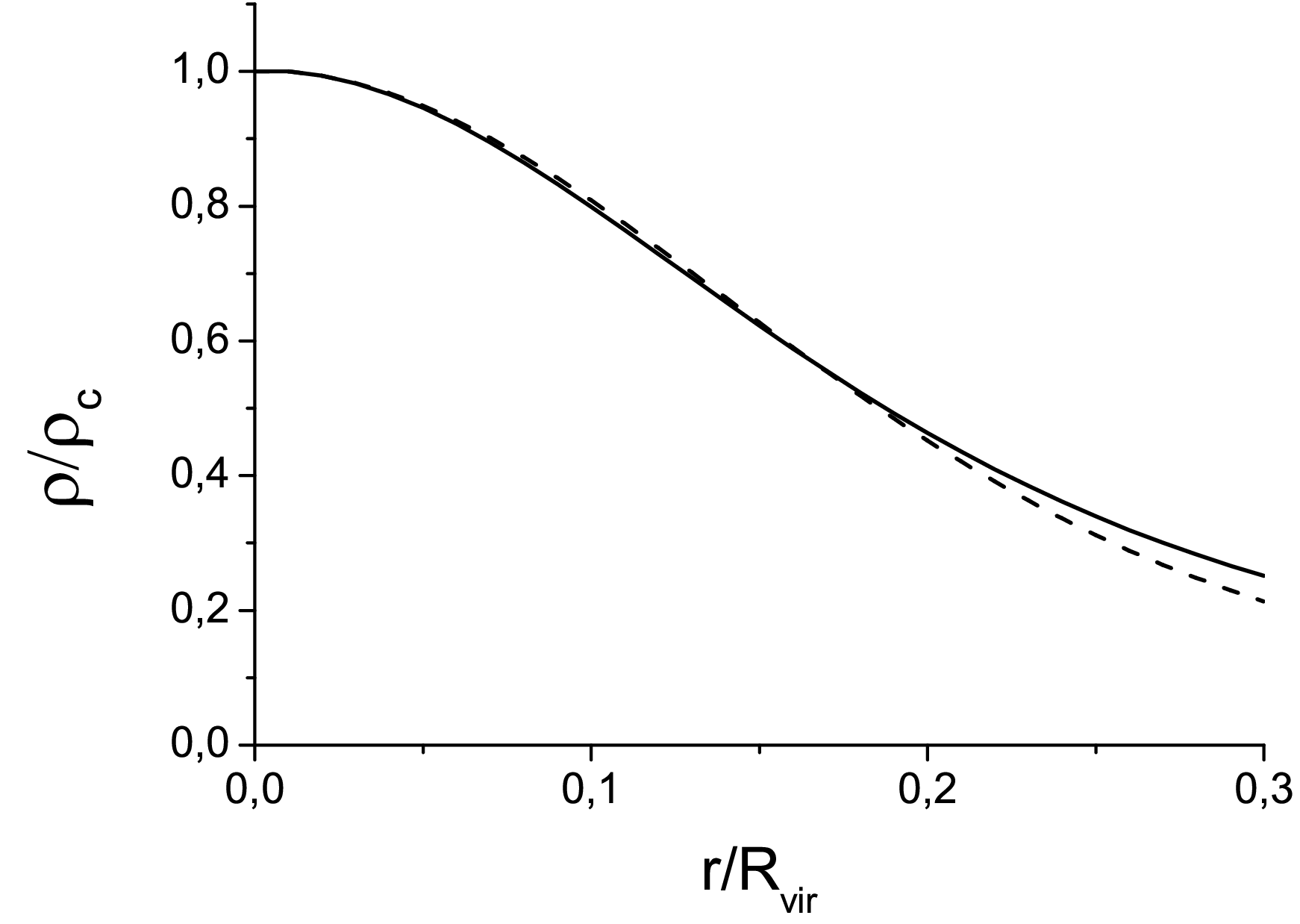}}
\caption{The density profile of the exact solution of (\ref{15a14}) for $f(r_0)=M_{vir}
\delta(r_0-R_{vir})$ (dashed line) and profile (\ref{15a20}) with the same $r_c/R_{vir}$ (solid
line). One can see that the departure of approximative equation (\ref{15a20}) from the exact
solution is quite small.}
 \label{15fig3}
\end{figure}

We can obtain the density distribution from (\ref{15a11}), substituting there (\ref{15a10}) instead
of $v_r$ and $4\pi r^2 \rho$ instead of $dm$:
 \begin{equation}
 \label{15a13}
\rho = \frac{m}{4 \pi r^2 v_r T} = \frac{m r_0}{4 \pi r T \sqrt{r_0^2-r^2}\:\sqrt{\mm^2-\mu^2} }
\end{equation}
In order to find the density distribution produced by all the particles of the halo with a certain
$r_0$, we should integrate (\ref{15a13}) over distribution (\ref{15a7}).
 \begin{equation}
 \label{15a14}
\rho = \frac{m r_0}{4 \pi r \sqrt{r_0^2-r^2}} \int\limits^{\mm}_0\!\frac{2 \mu
\exp\left(-\mu^2/\alpha^2\right)}{\alpha^2 T(r_0,\mu) \sqrt{\mm^2-\mu^2}}\; d\mu
\end{equation}
This is the exact solution; we can significantly simplify it, however, if we take into account that
$T(r_0,\mu)$ is in general a very weak function of $\mu$, especially for small $\mu$, since
$T(r_0,\mu)$ has an extremum at $\mu=0$. As we see in Fig.~\ref{15fig2}, $T(\mu)$ differs from
$T(0)$ on no more, than $25\%$, for any reasonable potential, if $\mu\le 0.5\mu_{circ}$, where
$\mu_{circ}$ is the angular momentum corresponding to the circular orbit. Only particles with small
$\mu$ can reach the central region under consideration: consequently, we may approximate $T$ by
$T(r_0,\mu)\simeq T(r_0)\equiv T(r_0,0)$: Then we can take the integral in (\ref{15a14})
 \begin{equation}
 \label{15a15}
\rho = \frac{m r_0}{2 \pi \alpha T(r_0) r \sqrt{r_0^2-r^2}}\; D\left(\dfrac{\mm}{\alpha}\right)
\end{equation}
where $D$ is the Dawson function $D(x)\equiv e^{-x^2}\int_0^x e^{t^2} dt$. Of course, all particles
in a real halo have some distribution over $r_0$: $dm=f(r_0) dr_0$. In order to take it into
account, we should substitute $f(r_0) dr_0$ instead of $m$ to (\ref{15a15}) and integrate the
result over $r_0$. Moreover, equation (\ref{15a15}) can be significantly simplified in the central
part of the halo. First of all, gravitational potential of the centre of the halo, $\phi(0)$, is
finite for any reasonable halo profile (otherwise the annihilation signal, produced by the halo,
would be infinite, being proportional to $\int 4\pi r^2 \rho^2 dr$). Then we may rewrite
(\ref{15a9}) as
 \begin{equation}
\label{15a17}
 \mm^2 = 2 r^2 (\phi(r_0)-\phi(0)) \left(1-\frac{\phi(r)-\phi(0)}{\phi(r_0)-\phi(0)}\right)
\frac{r^2_0}{r^2_0-r^2}
\end{equation}
The two last factors may be neglected in the center of the halo, and we can use approximations
$\mm\simeq r \sqrt{2 (\phi(r_0)-\phi(0))}$, $\sqrt{r_0^2-r^2}\simeq r_0$. So the total density
distribution in the central part is
 \begin{equation}
 \label{15a18}
\rho = \int^\infty_0\!\frac{f(r_0)}{2 \pi \alpha(r_0) T(r_0) r}\; D\left(r\dfrac{\sqrt{2
(\phi(r_0)-\phi(0))}}{\alpha(r_0)}\right) dr_0
\end{equation}
We can simplify this equation with the help of the following reasoning. As we could see, $f(r_0)$
is almost equal to zero for small $r_0$. Moreover, the formed halo is dominated by the particles
with $r_0\sim R_{vir}$. It means that the main contribution to the integral in (\ref{15a18}) is
given by the part, where $r_0\simeq R_{vir}$, roughly speaking, by $r_0\in [R_{vir}/2;R_{vir}$. As
we could see, function $f(r_0)$ sharply depends on $r_0$ at this interval. On the contrary,
$\alpha(r_0)$ should not change much on interval $[R_{vir}/2; R_{vir}]$: $\alpha(r_0)$ is widely
believed to be a power-law dependence with the index between $-1$ and $1$
\citep{2006JCAP...01..014H}. $\sqrt{2 (\phi(r_0)-\phi(0))}$ changes even slower: for instance,
$\sqrt{(\phi(R_{vir})-\phi(0))/(\phi(R_{vir}/2)-\phi(0))}\simeq 1.1$ for the NFW profile with
$c_{vir}=15$. Moreover, $D$ is a finite and not very sharp function if its argument. Comparing this
with the sharp behavior of $f(r_0)$, we may neglect the weak dependence of the argument of function
$D$ in (\ref{15a18}) on $r_0$ and substitute there some value, averaged over the halo
 \begin{equation}
 \label{15a19}
r_c = \left\langle\frac{\alpha(r_0)}{\sqrt{2
(\phi(r_0)-\phi(0))}}\right\rangle\simeq\frac{\langle\alpha(r_0)\rangle}{\sqrt{2 |\phi(0)|}}
\end{equation}
Then we can rewrite (\ref{15a18}) and obtain the final result:
 \begin{equation}
 \label{15a20}
\rho = \rho_c \frac{r_c}{r}\: D\!\left(\frac{r}{r_c}\right), \qquad \rho_c=\frac{1}{2 \pi
r_c}\int^\infty_0\!\frac{f(r_0) dr_0}{\alpha(r_0) T(r_0)}
\end{equation}
Since $D(r/r_c)\simeq r/r_c$, when $r/r_c \to 0$, $\rho_c$ is the central density of the halo. As
we can see, it is always finite. At the same time, the shape of the density profile depends on the
only parameter $r_c$. $r_c$ defines the radius at which rather a shallow central region of the
density profile transforms into a much steeper outer part. So is actually the core radius of
profile (\ref{15a20}).

In deriving (\ref{15a20}), we used several times the fact that $r\ll r_0$. This raises the
question: for what $r/R_{vir}$ is equation (\ref{15a20}) still valid? In order to check this, we
solved numerically exact equation (\ref{15a14}) on condition (\ref{15a7}) and for distribution
$f(r_0)=M_{vir} \delta(r_0-R_{vir})$. Since our approximation $T(r_0,\mu)\simeq T(r_0,0)$ is the
least acceptable for the particles with high angular momentum, we used the maximal possible value
for $\langle\alpha\rangle=\frac14 R_{vir} \sqrt{\frac{G M_{vir}}{R_{vir}}}=\frac14 \sqrt{G M_{vir}
R_{vir}}$. It is apparent that $\langle\alpha\rangle$ cannot substantially exceed this estimation:
otherwise a significant part of the dark matter would not be gravitationally bound in the halo.
$\phi(r)$ appears in expression (\ref{15a14}), and we need to write out equation
 $$
 \frac{d\phi}{dr}=\frac{G}{r^2}\int_0^r\!4\pi x^2\rho(x) dx
 $$
in order to close the task. The exact density profile (dashed line) and profile (\ref{15a20}) with
the same $r_c/R_{vir}$ (solid line) are represented in Fig.~\ref{15fig3}. We see that (\ref{15a20})
well approximates the exact solution up to $r\simeq 0.3 R_{vir}$. So (\ref{15a20}) can be valid in
quite a large area around the halo centre.

\begin{figure}
 \resizebox{\hsize}{!}{\includegraphics[angle=0]{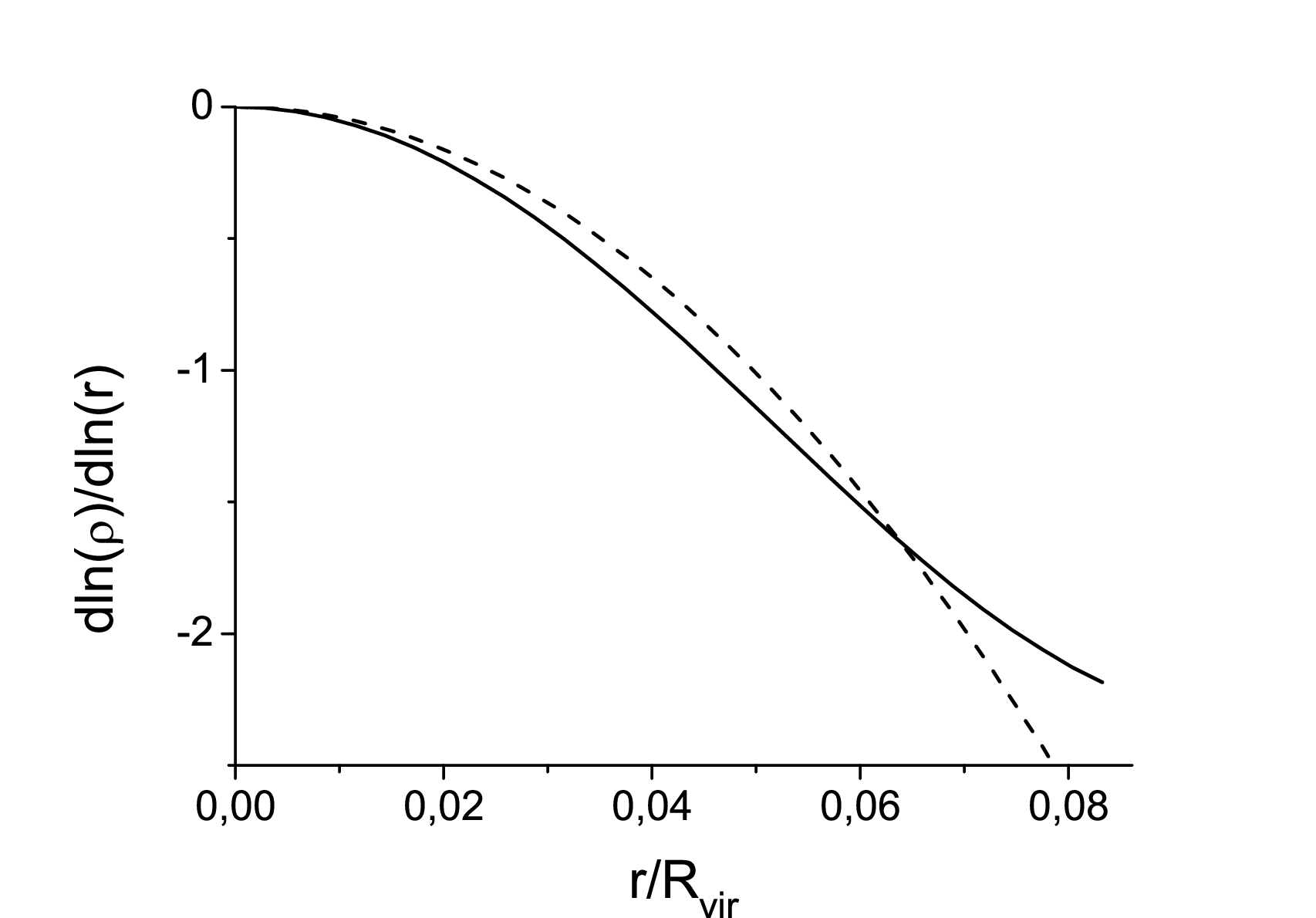}}
\caption{Density profile (\ref{15a20}) of the central region of a halo with $r_c=R_{vir}/20$.
Einasto profile with $n=0.5$ and $r_s=0.017 R_{vir}$ is plotted for comparison (dashed line).}
 \label{15fig4}
\end{figure}

\section{Discussion}
Fig.~\ref{15fig4} represents density profile (\ref{15a20}) for $r_c=R_{vir}/20$. For comparison,
Einasto profile with $n=0.5$ and $r_s=0.017 R_{vir}$ is also plotted. As we can see, the profiles
are quite similar, i.e. (\ref{15a20}) has a core and is close to Einasto profile with $n=0.5$,
which, in its turn, well fits observational dark matter profiles of numerous galaxies
\citep{mamon2011}.

There are two more properties of profile (\ref{15a20}) that can be observationally checked. First,
Fig.~\ref{15fig4} shows that (\ref{15a20}) behaves as $\rho\propto r^{-2}$ at large distances
($r\gtrsim 1.6 r_c$). The outer border of the profile (\ref{15a20}) applicability is defined by the
size of the area where distribution $f(r_0)$ is almost equal to zero (see the end part of section
\ref{sec2}), i.e., $r\simeq 2 R_{vir}/c_{vir}$. Since $1.6 r_c$ can be significantly smaller for
real galaxies, we may expect a large area with $\rho\propto r^{-2}$ in the density profile between
$r\gtrsim 1.6 r_c$ and $r\simeq 2 R_{vir}/c_{vir}$. This region is really observed in the profiles
of many galaxies \citep{rubinnew}. It is the feature that allowed to prove the existence of the
dark matter in 1978 \citep{rubinold}. Meanwhile, neither NFW nor Einasto profile has an explicit
region with $\rho\propto r^{-2}$. Of course, the power index $\frac{d\log\rho(r)}{d\log r}$ reaches
$-2$ at some radius. This point is nohow separated, however, and there is no reason to expect a
large region with the isothermal profile. Though there can be other reasons of its appearance, like
the baryon influence, the $\rho\propto r^{-2}$ profile persistently occurs in many galaxies with
very different properties, which implies a fundamental physical mechanism. Profile (\ref{15a20})
gives a simple and natural explanation of this effect.

We may roughly estimate the multiplication of the central density $\rho_c$ on the core radius $r_c$
\citep{16}. Indeed, if the particles with a certain $r_0$ have a characteristic velocity $v$, then
$\alpha(r_0)\sim r_0 v$, $T(r_0)\sim \tau_d =r_0/v$, and $\alpha(r_0) T(r_0)\sim r^2_0$.
Substituting this value into (\ref{15a20}), we obtain
\begin{equation}
 \label{15c2}
  \rho_c r_c\propto \int^\infty_0\!\frac{f(r_0) }{r^2_0} dr_0
\end{equation}
Now we can follow the same reasoning that we used to transform (\ref{15a18}) into (\ref{15a19}):
$f(r_0)$ is nonvanishing at $r\in [2 R_{vir}/c_{vir}, R_{vir}]$. At this interval $r^2_0$ varies
much slower than $f(r_0)$, and we may substitute it by some value $\langle r^2_0\rangle$ averaged
over the halo. Then
 \begin{equation}
 \label{15c3}
  \rho_c r_c\propto \dfrac{\int^\infty_0\! f(r_0) dr_0}{\langle r^2_0\rangle}=\dfrac{M_{vir}}{\langle r^2_0\rangle}
\end{equation}
As we could see, the majority of the particles have $r_0\sim R_{vir}$, and it is reasonable to
assume that $\langle r^2_0\rangle\propto R^2_{vir}$. So $\rho_c r_c\propto M_{vir}/R^2_{vir}$. It
is more convenient to use the average halo density $\varrho$: $M_{vir}=\frac43 \pi R^3_{vir}
\varrho$. We obtain
 \begin{equation}
 \label{15c4}
  \rho_c r_c\propto M^{1/3}_{vir} {\varrho}^{2/3}
\end{equation}
Thus our model predicts that the multiplication of the central halo density on the core radius
should be nearly constant for galaxies. Indeed, let us consider the haloes of masses $10^{8}
M_\odot$ and $10^{12} M_\odot$, which covers almost all the range of galaxy masses. We denote the
quantities related to $10^{8} M_\odot$ and $10^{12} M_\odot$ haloes by subindexes $8$ and $12$,
respectively. $M^{1/3}_{vir}$ increases only $\simeq 20$ times over all the mass interval under
consideration. Moreover, the second multiplier ${\varrho}^{2/3}$ decreases with $M_{vir}$, since
less massive galaxies formed earlier, when the universe density was higher. So the multiplier
variations partially compensate each other. Virial halo density ${\varrho}$ and $M_{vir}$ are not
one-to-one related. The structures appeared from a Gaussian random field, and objects of the same
$M_{vir}$ may have different $\varrho$. However, presuming that galaxies with $10^{8} M_\odot$ and
$10^{12} M_\odot$ were formed at $z_8=10$ and $z_{12}=4$, respectively, we may assume that their
density ratio is roughly equal to
$\frac{\varrho_8}{\varrho_{12}}\simeq\left(\frac{z_8+1}{z_{12}+1}\right)^3\simeq 8$
\citep{gorbrub2}. Thus our model predicts (\ref{15c4}) that $\rho_c r_c\propto M^{1/3}_{vir}
{\varrho}^{2/3}$ changes only $5$ times, when $M_{vir}$ runs over all the range of the galaxy
masses. This prediction is also strongly supported by observations. \citet{kormendy2004}  first
discovered the constancy of the multiplication of the central halo density on the core radius, and
then it was confirmed by several independent observations (see \citet{salucci2009} and references
therein). The observations suggest that $\log(\rho_c r_c)=1.88\pm 0.2$ in units of $\log (M_\odot
\text{pc}^{-2})$ for thousands of galaxies with various physical properties.

Thus the supposition of the moderate energy evolution along offers an explanation of three
observational properties of galactic haloes. The profile (\ref{15a20}) is in agreement with
observational data for, at least, some types of galaxies, but clearly disagrees with N-body
simulations and observational data for galaxy clusters \citep{okabe2010}. It suggests that the
violent relaxation takes place in galaxy clusters but does not occur in galaxies. There can be
various plausible explanations for it: baryon influence, warmth of the dark matter etc. However,
the cause may be much simpler: galaxy clusters have low concentrations $c_{vir}=3-5$. Condition
(\ref{15a1}) is very strict for an object with $c_{vir}=4$ (the energy of the majority of the
particles should not change on more that $30$\%) and unsatisfiable for $c_{vir}<3$. Galaxies have
significantly higher $c_{vir}$: $c_{vir}=15-20$ for massive galaxies and probably even higher for
dwarf objects. So in order to be violent, the relaxation should change the energies of a
significant part of the particles on $30-50$\% in the case of galaxy clusters and at least five
times in the case of galaxies. Consequently, the relaxation may be violent enough to form cusps in
galaxy clusters, but moderate in galaxies.

We can now specify the condition of feasibility of our \emph{principle assumption}: a moderate
energy evolution of the system during the halo formation. We are interested in the density profile
in the very center of the halo ($r\sim r_c$, where $r_c$ is the core radius). We can divide all the
particles in this area into two groups:
 \begin{enumerate}
 \item The particles residing this area (with $r_0\lessapprox r_c$)
\item The outer particles ($r_0\gg r_c$)
 \end{enumerate}
Both the groups give some yield into the density profile of the central region. Our consideration
(in particular, the derivation of  equation (\ref{15a20})) shows that the contribution of group 2
is always Einasto-like with small Einasto index ($n\simeq 0.5$). So, in order to ascertain that the
total density profile of the central part of the halo is close to (\ref{15a20}), we only need the
contribution of group 1 to be small. This requirement asserts automatically, if the energy exchange
between the dark matter particles is moderate in sense (\ref{15a1}).

We would like to summarize the main conclusions of this article. First, if the halo relaxation is
moderate, the density profile in the central part of a formed halo describes by equation
(\ref{15a20}), i.e. is Einasto-like with small Einasto index ($n\simeq 0.5$). The profile is quite
universal and weakly depends on the shape of the initial perturbation. This is a result of the fact
that the potential well of the formed halo is much deeper than that of the initial perturbation. As
a result, quite a large area builds up in the centre of the halo, where the particles do not
reside, but just come there from the outside: their apoapsides lie outside of the area. In order
that the orbit of a particle entirely lie within the area, the particle energy should be
dramatically decreased with respect to the initial value. Such a process is always hampered in a
dissipationless system, and hence the particles, coming to the central area from the outside,
dominate there and form Einasto-like universal density profile (\ref{15a20}). The shape of the
profile is totally determined by single parameter $r_c$ (\ref{15a19}), which can be considered as
the core radius. The model gives a natural explanation for the large areas of nearly isothermal
profile $\rho\propto r^{-2}$ routinely observed in spiral galaxies. Moreover, it predicts the
constancy of the multiplication of the central density $\rho_c$ on the core radius $r_c$ of dark
matter haloes, in good agreement with astronomical data.

\acknowledgments
 Financial support by Bundesministerium f\"ur Bildung und Forschung through
DESY-PT, grant 05A11IPA, is gratefully acknowledged. BMBF assumes no responsibility for the
contents of this publication. We acknowledge support by the Helmholtz Alliance for Astroparticle
Physics HAP funded by the Initiative and Networking Fund of the Helmholtz Association.

\end{document}